\documentclass{article}

\usepackage{amssymb,amsfonts,amsmath}
\usepackage{url}

\def\be{\begin{eqnarray}}
\def\ee{\end{eqnarray}}
\def\nn{\nonumber}

\def\tr{{\rm tr}\,}

\def\qD{\hbox{qD}}

\hoffset= - 1.0in         

\begin{document}
\title{\vspace{1.5cm}\bf\Large
Torus Knots in Adjoint Representation
}

\author{
Andrei Mironov$^{a,b,c,}$\footnote{mironov@lpi.ru,~mironov@itep.ru},~~
Vivek Kumar Singh$^{d,}$\footnote{vks2024@nyu.edu}
}

\date{ }

\maketitle

\vspace{-6cm}

\begin{center}
  \hfill FIAN/TD-23/25\\
  \hfill ITEP/TH-39/25\\
   \hfill IITP/TH-32/25
\end{center}

\vspace{4.5cm}

\begin{center}
 $^a$ {\small {\it Lebedev Physics Institute, Moscow 119991, Russia}}\\
$^b$ {\small {\it NRC ``Kurchatov Institute", 123182, Moscow, Russia}}\\
$^c$ {\small {\it Institute for Information Transmission Problems, Moscow 127994, Russia}}\\
$^d${\small{\it Center for Quantum and Topological Systems (CQTS), NYUAD Research Institute, \\
New York University Abu Dhabi, PO Box 129188, Abu Dhabi, UAE}}
\end{center}

\vspace{.1cm}

\begin{abstract}
We derive a closed-form expression for the adjoint  polynomials of torus knots and investigate their special properties.
The results are presented in the very explicit double sum form and provide a deeper insight into the structure of adjoint invariants essential for the Vogel's universality of Chern-Simons theory.
\end{abstract}

\bigskip

\newcommand\smallpar[1]{
  \noindent $\bullet$ \textbf{#1}
}

\paragraph{Introduction.}
Evaluating the HOMFLY-PT invariant of various knots and links is an important goal due to its numerous applications. Specifically, in accordance with \cite{Wit}, this invariant colored with a Young diagram $R$ is nothing but the Wilson loop average in representation $R$ in the $SU(N)$ Chern-Simons theory. Hence, its analysis reveals pure topological effects in Wilson averages of gauge theories \cite{Pol,Gua,Marino} as well as various Lie algebraic structures common for gauge theories\footnote{An example of a pure algebraic structure is given by the Vogel's universality, which can be conjecturally extended from Chern-Simons theory to Wilson averages in various gauge theories.}. In particular, the study of adjoint representation gives information about Wilson averages in the pure gluodynamics.

The adjoint invariants are also applied to study of the Vogel's universality of Chern-Simons theory \cite{Vog} (see also \cite{KhMS} for a latest review), which gives a unified description of various quantities in the adjoint  sector of representation theory for all simple Lie algebras in terms of three parameters $\alpha,\beta,\gamma$, which are homogeneous coordinates of the Vogel's projective plane. These quantities vary from the Chern-Simons partition function \cite{MkrtVes12} to the Racah matrix \cite{MMuniv} involving the adjoint representation and its descendants \cite{ManeIsaevKrivMkrt}. Notably, in \cite{MMM,BM}, one of us constructed universal invariants for the torus knots \(T[2,2n+1]\), \(T[3,3n\pm 1]\), and \(T[4,2n+1]\). Extending this construction to more general knots, however, has not yet been possible.

Moreover, there is also a pure knot theory application: the knowledge of the adjoint invariant allows one to evaluate the fundamental HOMFLY-PT polynomial of the reversed $2$-cable knot $K$,\footnote{An important point is that the invariants are here taken in the vertical framework \cite{MSSS,AM}.}
\be\label{cab}
H_{[1]}^{C_{2}(K)}(q,A) = \frac{1}{D_{[1]}(q,A)}\big(1 + D_{\mathrm{Adj}}(q,A)H_{\mathrm{Adj}}^{(K)}(q,A)\big),
\ee
and similarly for links \cite[Eq.(33)]{MSSS}, where $H_R$ denotes the (reduced) HOMFLY-PT invariant colored with the Young diagram $R$ and normalized to unknot, and $D_{[1]}$ ($D_{\mathrm{Adj}}$) is the quantum dimension of the fundamental (adjoint) representation.
In accordance with \cite[Eq.(2)]{MSSS}, this polynomial has a peculiar ``panhandle'' shape for torus knots, the case we deal with in this paper.

The torus knots \(T[m,n]\) are given by a pair of coprime numbers $m$ and $n$ (otherwise yielding a multi-component link), and are described as a contour put on a two-dimensional torus without self-intersections winding $n$ times around one cycle of the torus, and $m$ times around the other one (hence, $m\leftrightarrow n$ symmetry). Formula (\ref{cab}) for these knots becomes\footnote{The torus knot polynomial is here in the topological framing, hence, an additional factor of transition to the vertical framing $A^{-2(m-1)n}$ \cite{MSSS}.}
\be\label{cable}
H_{[1]}^{C_{2}(T[m,n])}(q,A) = \frac{1}{D_{[1]}(q,A)}\big(1 + A^{-2(m-1)n}D_{\mathrm{Adj}}(q,A)H_{\mathrm{Adj}}^{(T[m,n])}(q,A)\big),
\ee
and the panhandle shape of this polynomial means that, at $m<n$, it is a sum\footnote{Note that the condition $m<n$ is essential: restoring the symmetry of the torus knot $T(m,n)$ in $m$ and $n$ (topological framing) requires vanishing certain leading coefficients in Laurent polynomial $[1-2m; 2m-1]_A$ when $m>n$.}
\vspace{-2mm}
\[H_{[1]}^{C_{2}(T[m,n])}(q,A) =[1-2m; 2m-1]_A
\; +\;
\underbrace{(m-1) (q-q^{-1})
A\ \frac{
A^{2n}-
A^{2m}}{
A^{2}-1}}_{\rm{panhandle}}\,.\]
of quite a non-trivial ($n$-dependent) polynomial of $A$ and $q$ $[1-2m; 2m-1]_A$ of maximal degree $2m-1$ in $A$, and minimal degree $1-2m$, and of a simple panhandle polynomial of the degree growing with $n$.

Thus, we focus here on torus knots as a sufficiently large class of knots that admit a relatively simple evaluation of their invariants, and on the adjoint representation, which is, on the one hand, essential for various applications (the Vogel's universality, panhandle polynomials, etc.), and on the other hand, rather nontrivial, since the decomposition of its powers into irreducible representations contains nontrivial multiplicities. The evaluation of the HOMFLY-PT polynomial $H_R(q,A)$ is based on the Rosso-Jones formula \cite{RJ,LZ}, which, in turn, is based on the realization of $H_R(q,A)$ in terms of the ${\cal R}$-matrix of the algebra $U_q(sl_N)$ ($|q|<1$) in the representation $R$ \cite{RT}. Using the Rosso-Jones formula \cite{RJ,LZ} allows us to present the answer for the adjoint HOMFLY-PT polynomial of arbitrary torus knots.

\paragraph{Notation.} In this paper, we adopt the following notation:
\[
\{x\} := x - \frac{1}{x}, \qquad
[n]_q := \frac{\{q^n\}}{\{q\}}.
\]

\paragraph{Rosso-Jones formula.}
We start with the Rosso-Jones formula \cite{RJ,LZ} for the (normalized) invariant of the torus knot $T[m,n]$ ($m$ and $n$ are coprime) colored by a finite-dimensional irreducible representation $R$ of the simple Lie group associated with root system $\Delta$ ($\Delta_+$ denotes the positive roots):
\be
P^{[m,n]}_R(q,\Delta)\ = \ \frac{  q^{mn\varkappa_R}}{\qD_{\!_R}(q )} \sum_{Q\in R^{\otimes m}}
c_{_{RQ}}^{(m)} \cdot q^{-\frac{n}{m}\varkappa_{_Q}}\cdot \qD_{_Q}(q,\Delta).
\label{RJ}
\ee
Here
\be
\varkappa_Q = (\Lambda_Q,\Lambda_Q+2\rho)\nn
\ee
is the eigenvalue of the second Casimir operator in the representation $Q$, expressed in terms of its highest weight $\Lambda_Q$, and
\be\label{Weyl}
\qD_Q(q,\Delta)=\prod_{\alpha\in\Delta_+}{[\left(\Lambda_Q+\rho,\alpha\right)]_q\over [\left(\rho,\alpha\right)]_q}\nn
\ee
is the quantum dimension, and the coefficients $c_{_{RQ}}^{(m)}$ are defined by the Adams operation: $m$-plethysm
\be
\hat Ad_m \chi_{_R}(p_k) \equiv \chi_{_R}(p_{mk}) = \sum_{Q\in R^{\otimes m}}
c_{_{RQ}}^{(m)} \chi_{_Q}(p_k),
\label{Adrule}
\ee
where $\chi_{_R}(p_{mk})$ is the character of representation $Q$. In these formulas, $\rho={1\over 2}\sum_{\alpha\in\Delta_+}\alpha$ is the Weyl vector.
Note that expression (\ref{RJ}) is symmetric in $m$ and $n$ (which is absolutely non-trivial). The knot invariant (\ref{RJ}) is in the topological framing.

One of the most straightforward ways to evaluate the Adams coefficients is to use the explicit formula
\be
c_{_{RQ}}^{(m)}=\sum{\psi_R(\Delta)\psi_Q(m\Delta)\over z_\Delta},
\ee
where $\psi_R(\Delta)$ is the character of the representation $R$ of the permutation group on the conjugacy class given by the Young diagram $\Delta$, $z_\Delta$ is the order of automorphism of the Young diagram $\Delta$: if $\Delta=[\ldots,3^{r_3},2^{r_2},1^{r_1}]$, where some $r_k$ may be equal to zero, then $z_\Delta:=\prod_k k^{r_k}r_k!$. The Young diagram $m\Delta$ is understood as the Young diagram with all lengths of lines multiplied by $m$.

In the following, we specialize the general formula (\ref{RJ}) to the \( {\textbf{A}_{N-1}} \) series and derive explicit expressions for the adjoint HOMFLY-PT invariant of arbitrary torus knots.

\paragraph{${\textbf{A}_{N-1}}$ series. A simple example.}
For the \( {\textbf{A}_{N-1}} \) series, all ingredients entering the Rosso--Jones formula \eqref{RJ} admit a particularly explicit form. The finite-dimensional irreducible representation is labeled by the Young diagram $R$, and the corresponding character is given by the Schur function $S_R$ \cite{Mac}, which is realized either as a graded polynomial of $p_k=:\tr g^k$, where $g$ is the group element of $SL(N)$, or as a graded symmetric polynomial of the eigenvalues of $g$.

Consider the simplest example of the trefoil knot, which is $T[2,3]$ torus knot, and the fundamental representation, $R=\Box=[1]$. In this case, the character is
$S_{[1]}=p_1$ so that (\ref{Adrule}) becomes
\be
\hat Ad_2 S_{[1]}=p_2=S_{[2]}-S_{[1,1]}
\ee
the second Casimir eigenvalues are
\be
\varkappa_{[1]}=N-{1\over N},\ \ \ \ \ \varkappa_{[2]}=2+2N-{4\over N},\ \ \ \ \ \varkappa_{[1,1]}=-2+2N-{4\over N}
\ee
and the corresponding quantum dimensions are
\be
\qD_{[1]}={\{A\}\over\{q\}},\ \ \ \ \ \qD_{[2]}=\qD_{[1]}{\{qA\}\over\{q^2\}},\ \ \ \ \ \qD_{[2]}=\qD_{[1]}{\{A/q\}\over\{q^2\}}
\ee
Substituting all these quantities into (\ref{RJ}), one comes to
\be
H_{[1]}^{T[2,3]}=-{A^2\over q^2}\Big(q^2A^2-q^4-1\Big)
\ee
Note that this result can be equally well obtained with the standard skein relation (see the notation in \cite{skein})
\be
A^{-1}H_{[1]}(L_+)-AH_{[1]}(L_-)=(q-q^{-1})H_{[1]}(L_0)
\ee
which can be interpreted as a relation for the ${\cal R}$-matrix in the fundamental representation.
However, in the case of higher representations $R$, the skein relations do not work any longer \cite[sec2]{MMpol}, and one has to use instead the Reshetikhin-Turaev ${\cal R}$-matrix approach \cite{RT}, which ultimately reduces, in the case of torus knots, to the Rosso-Jones formula \cite{RJ}.

\paragraph{${\textbf{A}_{N-1}}$ series. Generic case.}
In the generic case in the ${\textbf{A}_{N-1}}$ series, the Casimir eigenvalue is
\be
\varkappa_Q = 2\sum_{\Box_{i,j}\in Q}(j-i)-{|Q|^2\over N}+|Q|N.
\ee

The quantum dimension is
\be
\qD_R(q,A)=\prod_{i,j\in R}{\{A q^{j-i}\}\over \{q^{h(i,j)}\}}~~,
\ee
where $h(i,j)$ is the hook length, and the product runs over boxes of the Young diagram.
It can be obtained by evaluating the Schur polynomial at the point $g=q^{2\rho}$ so that $\mathrm{tr} g^k={\{A^k\}\over\{q^k\}}$.

We will also make use of the notion of a \emph{composite representation} \cite{Koike}, which describes the most general finite-dimensional irreducible highest-weight representation of \( SL(N) \). Such a representation is associated with a Young diagram of the form
\vspace{-1mm}
$$
(R,P)= \Big[R_1+P_1,\ldots,R_{l_R}+P_1,\underbrace{P_1,\ldots,P_1}_{N-l_{\!_R}-l_{\!_P}},
P_1-P_{_{l_{\!_P}}},P_1-P_{{l_{\!_P}-1}},\ldots,P_1-P_2\Big],
$$
where $l_P$ denotes the number of lines in the Young diagram P. This $(R,P)$ is
the first (``maximal'') representation contributing to the product $R\otimes \bar P$, where $\bar P$ denotes the conjugate representation. It can be manifestly obtained from the tensor products (i.e., as a projector from  $R\otimes \bar P$) by formula \cite{Koike}
\vspace{3mm}
\be
(R,P)=\sum_{Y,Y_1,Y_2}(-1)^{l_{\!_Y}}N^R_{YY_1}N^{P}_{Y^TY_2}\ Y_1\otimes\overline{Y_2}\,,
\ee
 where the superscript ``T'' denotes transposition.

The adjoint representation is the simplest example of the composite representation, it is given by the Young diagram $[21^{N-2}]$, which depends on $N$. We introduce the variable $A:=q^N$ as a convenient notation for our discussion. This variable $A$ should not be confused with the second variable $A$ of the HOMFLY-PT polynomials. The HOMFLY-PT polynomial is evaluated for one and the same Young diagram for various ${\textbf{A}_{N-1}}$, while the adjoint (or, more generally, composite) representation depends on $N$ itself. Formally, we define $H_{\mathrm{Adj}}(q,A)$ via the specialization
\be
H_{\mathrm{Adj}}(q,A) \;\big|_{\,A=q^N} \;=\; J_{[2,1^{N-2}]}^{SL(N)}(q),
\ee
for all sufficiently large $N$ \cite{ChE}\cite{HM}. Here $J_R^{SL(N)}(q)$ is the $SL(N)$ invariant. This polynomial is called \emph{uniform HOMFLY-PT polynomial} in \cite{MMM}, and the same construction applies to any composite representation.

Our goal now is to construct the uniform (adjoint) HOMFLY-PT polynomial for the torus knots $T[m,n]$, i.e. $m$ and $n$ are coprime.
Note that, in this case, constructing torus knot invariants is generally simpler than links: for instance, $\mathrm{Adj}^{\otimes 4}$ decomposes into 40 irreps, while the Adams operation generates only 17 irreps \cite{BM}. This is because the plethystic expansion of the adjoint Schur function generating the Adams coefficients is relatively simple. For instance, the 5-plethystic expansion is
\be\label{A5}
\hat Ad_5 S_{\mathrm{Adj}}&=&4+S_{[2^51^{N-10}]}-S_{[3^52^{N-9}1^3]}+S_{[4^53^{N-8}2^2]}-S_{[5^54^{N-7}3]}+S_{[6^55^{N-6}]}-\nn\\
&-&S_{[32^31^{N-9}]}+S_{[43^32^{N-8}1^3]}
-S_{[54^33^{N-7}2^2]}+S_{[65^34^{N-6}3]}-S_{[76^35^{N-5}]}+\nn\\ &+&
S_{[42^23^{N-8}]}-S_{[53^22^{N-7}1^3]}+S_{[64^23^{N-6}2^2]}
-S_{[75^24^{N-5}3]}+S_{[86^25^{N-4}]}-\nn\\
&-&S_{[521^{N-7}]}+S_{[632^{N-6}1^3]}-S_{[743^{N-5}22]}+S_{[854^{N-4}3]}-S_{[965^{N-3}]}+\nn\\
&+&S_{[61^{N-6}]}
-S_{[72^{N-5}1^3]}+S_{[83^{N-4}22]}-S_{[94^{N-3}3]}+S_{[10,5^{N-2}]}~,
\ee
which has to be compared with just antisymmetric part of $\mathrm{Adj}^{\otimes 5}$ in \cite{IK24}.

For the adjoint representation, the quantum dimension and the Casimir exponential take particularly simple forms:
\be\label{qDA}
\begin{array}{rclrcl}
\qD_{\mathrm{Adj}}(q,A)=\qD_{[21^{N-2}]}(q,A)&=&\displaystyle{\{Aq\}\{A/q\}\over\{q\}^2},~~
\hspace{1cm}&q^{\varkappa_{\mathrm{Adj}}} &=&A^2.\cr
\end{array}
\ee

\paragraph{Uniform polynomials for torus knots $T[m,n]$.}
Because of simplicity of the plethystic expansion, it is technically available to construct the knot invariant in the adjoint representation for the torus knots $T[m,n]$ with arbitrary coptime $m$ and $n$. The Adams operation generates $m^2+1$ irreps in this case. The general answer is quite transparent:
\be
\hat Ad_m S_{\mathrm{Adj}}&=&m-1+\nn\\
&+&\underbrace{S_{[2^m1^{N-2m}]}-S_{[3^m2^{N-2m+1}1^{m-2}]}
+S_{[4^m3^{N-2m+2}2^{m-3}]}+...+(-1)^{m+1}S_{[m+1,m^{N-m-1}]}}_{m\ terms}-\nn\\
&-&\underbrace{S_{[32^{m-2}1^{N-2m+1}]}+\ldots}_{m\ terms}+\nn\\
&+&\underbrace{(-1)^mS_{[42^{m-3}1^{N-2m+2}]}+\ldots}_{m\ terms}+\nn\\
&\ldots&\nn\\
&+&\underbrace{(-1)^{m+1}S_{[m+1,1^{N-m-1}]}+(-1)^mS_{[m+2,2^{N-m}1^{m-2}]}+\ldots+ S_{[2m,m^{N-2}]}}_{m\ terms},
\ee
with totally $m^2+1$ terms in this formula.

With these formulas for the Adams operation, constructing the adjoint knot invariants is immediate. To this end, we rewrite these formulas in the composite representation notation and notice that the Adams operation gives rise to only hook Young diagrams:
\be\label{14}
\hat Ad_m S_{\mathrm{Adj}}=m-1+\sum_{a,b=1}^m(-1)^{a+b}S_{([a,1^{m-a}],[b,1^{m-b}])}.
\ee
In order to justify this formula, one can note that the coefficients $c^{(m)}_{RQ}$ in the Adams operation can be obtained by splitting the decomposition of $R^{\otimes m}$ into a sum of terms with fixed symmetric patterns given by the Young diagrams $P$ of size $m$:
\be
R^{\otimes m}=\sum_{P\vdash m}\pi_P\left(R^{\otimes m}\right).
\ee
Then, the Adams operation gives (see \cite{RJ},\cite{LZ})
\be\label{Ad}
\hat Ad_m (R)=\sum_{P\vdash m}\psi_P([m])\pi_P\left(R^{\otimes m}\right),
\ee
where $\psi_P([m])$ is the value of the character of the permutation group ${\cal S}_m$ in the representation $P$ on the cycle of the maximal length (i.e. on the cyclic permutation $(1,\ldots,m)$):
\be
\psi_P([m])=\left\{\begin{array}{cc}
(-1)^d&\hbox{if }P=[m-d,1^d]\\
0&\hbox{otherwise}.
\end{array}\right.
\ee

Now we evaluate the second Casimir eigenvalues of the representations entering (\ref{14}). For the composite representation $(R,P)$ with $|R|=|P|=m$, it is generally given by the formula (see, e.g., \cite{MSSS})\footnote{In the generic case, the formula has the form
$$
\kappa_{(R,P)}=2N(|R|+|P|)-{(|R|-|P|)^2\over N}+\kappa_R-\kappa_{P^T}.
$$}
\begin{eqnarray}\label{p}
\kappa_{(R,P)}&=&2Nm+\kappa_R-\kappa_{P^T}\nn\,.
\end{eqnarray}
 Using
\be
\kappa_{[a,1^b]}=a(a-1)-b(b+1),\nn
\ee
one obtains
\be
q^{\kappa_{([a,1^{m-a}],[b,1^{m-b}])}}&=&A^{2m}q^{2m(a+b-m-1)}.
\ee
Similarly, the quantum dimension of the composite representation is given by the formula \cite{MM}, which can be rewritten in the form
$$
\qD_{(R,P)}(q,A)
=  \qD_{_R}(q,Aq^{-l_{\!_P}})\, \qD_{_P}(q,Aq^{-l_{\!_R}})\,
\prod_{i=1}^{l_{\!_R}}\prod_{j=1}^{l_{\!_P}}
{[N+R_i+P_{j}+1-i-j]_q\over[N+1-i-j]_q},
$$
so that
\be\label{qD1}
\qD_{([a,1^{m-a}],[b,1^{m-b}])}(q,A)={\{Aq^{a+b-1}\}\{Aq^{a-m-1}\}\{Aq^{b-m-1}\}\over
\{q^m\}^2\{Aq^{-1}\}}\
\prod_{i=1}^{a-1}{\{Aq^{i+b-m-1}\}\over \{q^i\}}
\prod_{i=1}^{b-1}{\{Aq^{i+a-m-1}\}\over\{q^i\}}
\times\nn\\
\times \prod_{i=1}^{m-a}{\{Aq^{-i+b-m-1}\}\{Aq^{b-i}\}
\over\{q^i\}\{Aq^{-1-i}\}}
\prod_{i=1}^{m-b}{\{Aq^{-i+a-m-1}\}\{Aq^{a-i}\}
\over\{q^i\}\{Aq^{-1-i}\}}
\prod_{i=1}^{m-a}\prod_{j=1}^{m-b}{\{Aq^{1-i-j}\}\over\{Aq^{-1-i-j}\}},
\ee
since
$$
\qD_{([a,1^{m-a}])}(q,A)={\prod_{i=0}^{a-1}\{Aq^i\}\prod_{i=1}^{m-a}\{Aq^{-i}\}\over
\{q^m\}\prod_{i=1}^{a-1}\{q^i\}\prod_{i=1}^{m-a}\{q^i\}}.
$$

Thus, we finally come to the main result of this paper, to the explicit formula for the (reduced or normalized) uniform HOMFLY-PT polynomial of the torus knot $T[m,n]$:
\be\label{main}
\boxed{
H^{T[m,n]}_{\mathrm{Adj}}(q,A)\ = \ \frac{  A^{2mn}}{\qD_{\mathrm{Adj}}(q )}\left(m-1+ \sum_{a,b=1}^m
(-1)^{a+b} A^{-2n}q^{-2n(a+b-m-1)}\cdot \qD_{([a,1^{m-a}],[b,1^{m-b}])}(q,A)\right),
}
\ee
with the quantum dimensions given by formulas (\ref{qDA}) and (\ref{qD1}).

Substituting the result \eqref{main} into \eqref{cable}, we immediately obtain the fundamental normalized HOMFLY-PT polynomial of the reverse $2$-cable $T(m,n)$ knot \cite{MSSS}. The following section is devoted to a detailed study of the properties of the obtained formulas\eqref{main}.

\paragraph{Properties of uniform polynomials.}

Using the formulas (\ref{main}), one can verify for a generic torus knot \( T(m,n) \) that the following properties hold:
\begin{itemize}
\item As is usual with the Rosso-Jones formula, it looks non-symmetric in $n$ and $m$ for the torus knot $T[n,m]$, but, in fact, it is symmetric. In particular, this means that one can choose the smaller of two numbers $m$ and $n$ as an upper limit of summation in (\ref{main}) to have a simpler sum.
\item Since $T[m,1]$ is unknot,
\be
H^{T[m,1]}_{\mathrm{Adj}}=1.
\ee
\item
In the case of $T[m,0]$, which is the pure plethysm,
\be
H^{T[m,0]}_{\mathrm{Adj}}={\qD_{\mathrm{Adj}}(q^m,A^m)\over\qD_{\mathrm{Adj}}(q,A)}.
\ee
\item One checks that the stable {\bf uniform} HOMFLY-PT homology of torus knots (in the spirit of \cite{GGS}) is
\be
\lim_{n \to \infty }\left({q\over A}\right)^{2n(m-1)}{\cal H}^{T[m,n]}_{\mathrm{Adj}}=\qD_{([m],[m])}(q,A)=
\qD_{m \mathrm{Adj}}(q,A),
\ee
where ${\cal H}^{T[m,n]}_{\mathrm{Adj}}$ denotes the unreduced uniform HOMFLY-PT polynomial. This immediately follows from (\ref{main}) at $|q|<1$, $|A|<1$.

\item
One can also check that
\be
H^{T[m,n]}_{\mathrm{Adj}}\Big|_{q=e^\hbar}=1+O(\hbar^2),
\ee
which just means that the framing is topological.
\end{itemize}
Particular specializations of the polynomial $H^{T[m,n]}_{\mathrm{Adj}}$ are:
\begin{itemize}
    \item
{\bf The Alexander polynomial}
\be
\mathfrak{A}^{T[m,n]}_{\mathrm{Adj}}=H^{T[m,n]}_{\mathrm{Adj}}\Big|_{A=1}=1.
\ee
\item {\bf The special polynomial}
\[
\sigma^{T[m,n]}_{\mathrm{Adj}}(A) := H^{T[m,n]}_{\mathrm{Adj}}\Big|_{q=1},
\]
which, for the series of torus knots $T[2,n=2k+1]$, $T[3,n=3k\pm 1]$, $T[4,n=2k+1]$, $\ldots$, is equal to (the signs in front of $\sigma_{[1]}^{T[m,n]}$ are, of course, not seen in the uniform special polynomial)
\be
\sigma^{T[m,n]}_{\mathrm{Adj}}(A)&=&\left(\sigma_{[1]}^{[m,n]}\right)^2\nn,\\
\sigma_{[1]}^{T[2,n=2k+1]}&=&-{1\over 2}A^{n-1}\Big((n-1)A^2-(n+1)\Big)\nn,\\
\sigma_{[1]}^{T[3,n=3k\pm 1]}&=&{1\over 6}A^{2(n-1)}\Big((n-1)(n-2)A^4-2(n-1)(n+1)A^2+(n+1)(n+2)\Big)\nn,\\
\sigma_{[1]}^{T[4,n=2k+1]}&=&-{1\over 24}A^{3(n-1)}\Big((n-1)(n-2)(n-3)A^6-3(n-1)(n-2)(n+1)A^4+\nn\\
&+&3(n-1)(n+1)(n+2)A^2-(n+1)(n+2)(n+3)\Big)\nn,\\
&\ldots&
\ee
This, indeed, coincides with the special polynomial of the torus knot $T[m,n]$ with arbitrary coprime $m$ and $n$ in the fundamental representation:
\be
\boxed{
\sigma_{[1]}^{T[m,n]}=
(-1)^{m+1}{1\over m}A^{(m-1)(n-1)}\cdot
\sum_{k=1}^m(-1)^{k+1}\binom{n-1}{m-k}\binom{n+k-1}{k-1}A^{2(m-k)}
.}
\ee
The symmetry of this expression in $m$ and $n$ is absolutely non-evident, but it is symmetric.
\item The $A=q^2$ corresponds to the spin 1 representation of $SL(2)$, and, hence the adjoint HOMFLY-PT polynomial becomes {\bf the colored Jones polynomial}: $J^{T[m,n]}_2(q):=H^{T[m,n]}_{\mathrm{Adj}}\Big|_{A=q^2}$. In fact:
\be
\boxed{
H^{T[m,n]}_{\mathrm{Adj}}\Big|_{A=q^2}={q^{2(m-1)(n-1)}\over 1-q^6}\left(1-q^{4m+2}-q^{4n+2}+q^{4(n+m)}
+q^{2(m+1)(n+1)-2}(1-q^2)\right),
}
\ee
which coincides with the colored Jones polynomial $J_2$ of the torus knot $T[m,n]$. In this case, the symmetry between $m$ and $n$ is manifest.
\end{itemize}

\paragraph{Concluding remarks.}
We have presented an explicit formula (\ref{main}) for the adjoint HOMFLY-PT polynomial for an arbitrary torus knot $T[m,n]$. It is quite surprising to find a closed-form formula for the HOMFLY-PT polynomial in a mixed representation: until now, this was only possible for polynomials in (anti)symmetric (or, possibly, rectangular) representations (i.e., those whose square decomposition does not contain multiplicities). Importantly, such an explicit form is necessary for constructing universal Vogel's formulas (see \cite{BM} for the last example).

A natural extension of the present analysis is to the case of torus links, which is considerably more involved. The complexity arises for two main reasons. First of all, the number of representations $Q$ that contribute to the r.h.s. of the Rosso-Jones formula in the case of knots is typically less than half that in the case of links. This is because the $m$-plethysm expansion of the adjoint representation gives rise to composite representations $(R,P)$ with $|R|=|P|=m$ only (and not all of possible representations even with this restriction emerge). In the case of $l$-component links, the sum in the Rosso-Jones formula runs over representations with $|R|=|P|$ equal to $mk/l$ with $k=1,\ldots,l$ (see \cite{MSSS}). An additional complication arises from the non-trivial multiplicities of representations appearing in the sum. Due to these chain of reasoning, calculating the adjoint HOMFLY-PT invariants for arbitrary links remains a challenging problem. It therefore represents an important topic for future investigation.

Another essential problem that remains is to make a similar calculation for arbitrary knots, but for other groups. This would allow one to construct the universal invariant in the Vogel's sense. Finally, it would be interesting to generalize the present construction to refined Chern--Simons theory, in the spirit of \cite{BMM}.

\paragraph{Acknowledgements.}

We are grateful to Alexander Stoimenov and Hisham Sati, whose intelligent questions were one of the reasons that prompted us to conduct this research. The work of A.M. was partially funded within the state assignment of the NRC Kurchatov Institute, was partly supported by the grant of the Foundation for the Advancement of Theoretical Physics and Mathematics “BASIS” and by Armenian SCS grants 24WS-1C031. The work of V.K.S. is supported by Tamkeen UAE under the
NYU Abu Dhabi Research Institute grant {\tt CG008}.

\end{document}